# Observation of a spin-textured nematic Kondo lattice

**Authors:** Yu-Xiao Jiang[1]†*, Zi-Jia Cheng[1]*, Qiaozhi Xu[4]*, Md Shafayat Hossain[1], Xian P. Yang[1], Jia-Xin Yin[5], Maksim Litskevich[1], Tyler A. Cochran[1], Byunghoon Kim[1], Eduardo Miranda[3], Sheng Ran[4], Rafael M. Fernandes[2,**], M. Zahid Hasan[1]†

**Affiliations:**

[1]Laboratory for Topological Quantum Matter and Advanced Spectroscopy, Department of Physics, Princeton University, Princeton, New Jersey, USA.

[2]School of Physics and Astronomy, University of Minnesota, Minneapolis, MN 55455, USA.

[3]Gleb Wataghin Physics Institute, The University of Campinas, 13083-859, Campinas, SP, Brazil.

[4]Department of Physics, Washington University in St. Louis, St. Louis, MO 63130, USA

[5]Department of physics, Southern University of Science and Technology, Shenzhen, Guangdong 518055, China.

†Corresponding authors, E-mail: yuxiaoj@princeton.edu; mzhasan@princeton.edu

*These authors contributed equally to this work.

**Present address: Department of Physics and Anthony J. Leggett Institute for Condensed Matter Theory, The Grainger College of Engineering, University of Illinois Urbana-Champaign, Urbana, IL 61801, USA

**Abstract:**

**The Kondo lattice model, as one of the most fundamental models in condensed matter physics, has been employed to describe a wide range of quantum materials such as heavy fermions[1-4], transition metal dichalcogenides and two-dimensional Moire systems[14-17]. Discovering new phases on Kondo lattice and unveiling their mechanisms are crucial to the understanding of strongly correlated systems. Here, in a layered Kondo magnet USbTe, we observe a spin-textured nematic state and visualize a heavy electronic liquid-crystal phase. Employing scanning tunneling microscopy and spectroscopy (STM/STS), we visualize a tetragonal symmetry breaking of heavy electronic states around the Fermi level. Through systematically investigating the temperature and energy dependence of spectroscopic data, we find that the nematic state coincides with the formation of heavy quasi-particles driven by band hybridization. Remarkably, using spin polarized STM, we demonstrate that the nematic state is spin polarized, which not only suggests its intrinsically electronic nature, but also represents the unique magnetic texture of nematic heavy fermions. Our findings unveil a novel correlation-mediated order whose mechanism is inherently tied to Kondo-lattice physics. The observation of heavy nematic states enriches the phase diagram of correlated systems and provides a rare platform to explore the interplay of Kondo physics[3-4], spontaneous symmetry breaking[12,13] and quantum criticality[9-11].**

**Main Text:**

In a Kondo lattice, the interaction between local moments and itinerant electrons is essential to the emergence of exotic states of matter[1-4]. As temperature cools down, the *f*-orbital electrons, which are usually localized in real-space, interact with conventional nearly free electrons and hybridize with them, becoming part of the Fermi Sea and endowing it with heavy effective masses. Over the past decades, a tremendous number of quantum phenomena have been observed in heavy fermion systems such as unconventional superconductivity[18-25], non-Fermi liquid[11,26-29] and exotic orders[30]. On the other hand, the nematic electronic liquid-crystal phase is an intriguing order often observed in high-temperature superconductors and, in some cases, associated with non-Fermi liquid behavior[6]. Although evidence of rotational symmetry breaking states has been observed in heavy fermion materials via macroscopic probes[31-35], its microscopic connection to heavy fermions and Kondo physics is largely unknown. In this work, we report the observation of a heavy electronic liquid-crystal phase in a Kondo magnet USbTe. Combining STM, ARPES and theoretical analysis, we show that the heavy electronic liquid-crystal phase: 1) breaks four-fold rotational symmetry 2) consists of heavy quasi-particles 3) is spin polarized and therefore distinct from existing phases of matter[31-36,46-48].

USbTe is a heavy fermion ferromagnet with Curie temperature $T_c$ = 125K [37-41]. It has a layered structure, consisting of planes of Uranium (U) and Tellurium (Te) atoms sandwiched by Antimony (Sb) planes. Within each plane, the corresponding atoms form a square net preserving the mirror symmetry and the $C_4$ rotational symmetry of the crystal lattice (corresponding to a 90-degree rotation with respect to the c-axis). In our STM experiments, crystals typically cleave between the U and Sb planes due to their relatively weak chemical bonding and show the (001) surface, as shown in Fig. 1**a**. After extensive searches in 4 samples, we have found two kinds of terminations A and B, as shown in Fig. 1**b-c**. The arrangement of individual atoms is marked in the two terminations, each forming a square lattice. Since the distance between U-U atom is $\sqrt{2}$ larger than Sb-Sb atom, we attribute termination A to the Sb plane and termination B to the U plane. The identification of U termination is further supported by spectroscopic data and spin-polarized measurements, which we will present in the following. Fig. 1**d** demonstrates a typical topography taken on the U plane (termination B), featuring a flat surface with random atomic defects. The atomically resolved topography shows the high quality of the surface. The exposure of U atoms allows us to directly probe the heavy fermion behavior. In the remainder of this paper, the data are all taken on the U termination unless otherwise noted.

In a heavy fermion lattice, the localized moments from the f orbitals (which here come from the U atoms) are immersed in the itinerant electron sea (which here can come from d-orbitals of the U atoms or p-orbitals of the Sb atoms). When electrons tunnel into the system, the tunneling conductance resembles the shape of a Fano spectrum, known as Kondo resonance[42]. In STM experiments, the tunneling junction is built up between the STM tip and the sample, allowing us to access the local differential tunnelling conductance dI/dV(V). Figure 1e shows a spatially averaged tunneling spectrum taken at 20K, exhibiting features of Kondo resonance. The conductance can be fitted by a Fano spectrum parameterized by

$$\mathrm{dI/dV}(V) \propto \frac{[(V - E_0)/\Gamma + q]^2}{1 + (V - E_0)/\Gamma^2}$$

where $E_0$ is the resonance energy, $\Gamma$ is the resonance width and $q$ is the Fano parameter. Upon cooling the sample below 15K, an asymmetric peak-dip feature develops near the Fermi level (marked by the black arrows). The peak-dip feature of the dI/dV curves has been observed on U terminations of other heavy fermion compounds[49-52]. It has been suggested that the enhanced spectral weight near the Fermi level can come from the heavy f-electrons, and the spectrum dip is related to the partially opened hybridization gap[52]. Remarkably, transport measurement of the bulk sample shows a resistivity upturn below 15K[37,39], concurrent with the peak-dip feature we observed.

We present the main finding of the work, a nematic Kondo lattice, in Figure 2. The STM differential conductance map dI/dV(V, x) provides the energy resolved electronic density of states distribution. Figure 2**a** shows the real space dI/dV conductance map at the Fermi energy ($E_f$), i.e. V = 0 mV, taken at 4.2K. The conductance map exhibits a stripe-like pattern, whose direction is oriented 45 degrees with respect to both *a* and *b* axis, as marked by the yellow arrow in Fig. 2**a**. The stripe pattern is reminiscent of the fingerprints of electronic nematicity seen in iron-based superconductors[44,47,48], breaking the $C_4$ rotational symmetry of the crystal lattice, and we thus attribute it to a nematic state. Notably, the dashed line in Fig.2**a** tracks a boundary between two domains, where the unidirectional patterns are rotated by 90 degrees. The observation of multiple domains in the same field of view can rule out anisotropic STM tip effects. Extended Fig. 1**a-b** shows a large-area topographic image and its Fourier transform. The flat surface and isotropic Bragg peaks demonstrate the lack of structural disorder. Figure 2**c** displays representative energy-resolved dI/dV maps focusing on the same region. While the stripe pattern is prominent at $E_f \pm 2$ meV, it almost disappears at $E_f \pm 25$ meV. The energy dependence of the dI/dV maps illustrates that the nematic state mainly emerges at energies close to the Fermi level. The anisotropy direction of the nematic state is pinned in each domain regardless of atomic defects, suggesting it does not arise from an anisotropic scattering potential.

We further investigate the temperature dependence of the nematic state. Figure 2**d** shows a series of spectroscopic maps dI/dV(x) at E = 10 meV, close to the Fermi surface. The real space stripe pattern, evident at 4.2K, is substantially weakened at 15K. By examining the Fourier-transformed (FT) dI/dV maps at different temperatures (Extended Fig.10), we observe that the $C_4$ symmetry is broken at 4.2K and restored as temperature rises above 15K. To quantitatively extract the anisotropy of the electronic state, we define a factor $\alpha = Tr[(M - M^R)(M - M^R)^T]$, where $M$ is the matrix for normalized FT images and $M^R$ is the same matrix but rotated by 90 degrees (details provided in Methods). For a perfectly $C_4$ symmetric FT image, $\alpha$ should be 0, while the breaking of $C_4$ symmetry will lead to a non-zero $\alpha$ value. In practice, however, due to noise inevitably present in the experiments, $\alpha$ is not identically zero even in a nominally tetragonal sample. We calculate $\alpha$ for the FT dI/dV maps at different energy cuts and temperature, and the results are presented in Fig. 2**e**. We make two important observations: First, at 4.2K, $\alpha$ peaks around the Fermi level, decreases as the energy becomes higher (or lower), and eventually fluctuates at the noise background level at 20 meV away from the Fermi level. Second, at 15K, the peak around the Fermi level is gone and the energy dependence of $\alpha$ becomes essentially featureless. Remarkably, the nematic state exhibits similar energy and temperature scales as the dI/dV feature. Below 15K, the nematicity occurs at low energy where the dI/dV peak-dip is located.

We now show the rotational symmetry breaking of heavy quasi-particles using spectroscopic tools. In STM measurements, the scattering process between quasi-particles and defects can produce signals in the FT dI/dV maps, enabling us to infer the energy and momentum structure of the electronic bands, known as quasi-particle interference (QPI) maps. Figure 3**a** shows a series of QPI maps at selected energies, which are two-fold symmetrized to enhance the signal to noise ratio (raw data in Extended Fig.2). The Bragg vectors $Q_{a,b}$ and other reference points are labeled. Moving toward the Fermi level, the QPI map evolves from a $C_4$ to a $C_2$ symmetric pattern, as a manifestation of nematic order. Along $OM'$, we observe sets of scattering vectors $q$, marked by the green arrow at -5 meV. The corresponding scattering intensity surrounds the origin and roughly forms a square shape. The QPI band dispersion suggests that $q$ comes from scattering of a light conduction band. Figure 3**b** shows the line cut of the QPI intensity along the $OM$ and $OM'$ directions with finer energy resolution. In the QPI dispersion along $OM$, the vector $q$ shows light band behavior with a steep slope, marked by the white dot in Fig. 3**b**. However, along the 90-degree rotated direction ($OM'$), the QPI band changes dramatically: the slope of the QPI band becomes strongly energy dependent, and a flat intensity appears around 5 meV, indicating the formation of heavy quasi-particles[49-52]. Importantly, the flat intensity breaks the four-fold rotational symmetry and only appears along one of the two orthogonal cuts ($OM$ and $OM'$). By contrast, at 22K, the light band persists through the whole energy window along the two orthogonal cuts (Fig. 3**c**). To validate the existence of a flat band at the Fermi level, we have also conducted ARPES measurements. Figures 3**d,e** shows the ARPES intensity cut along $M - \Gamma - M$ at 6 K and 130 K. The white arrow points to a steep light band. At 6K, we observe intensity (purple) forming a flat shape at the Fermi level and hybridize with the conduction band. Figure 3**f** shows the Fermi surface, where the Brillouin zone and high-symmetry momentum points are labeled. The flat band feature exhibits intensity (the blue region) dominantly along $\Gamma M$ (compared to $\Gamma X$), consistent with the QPI signal. However, we note though our ARPES data locates the flat band along $\Gamma M$, the pattern in figure 3**f** cannot show the real symmetry of Fermi surface. The asymmetry in the ARPES Fermi surface map typically comes from the photoemission matrix element effects and experimental set-up. The QPI and ARPES measurements demonstrate that, around the Fermi level, the conduction band is hybridized with the flat band from the *f*-orbitals, leading to heavy quasiparticles. Concurrently, the QPI signals exhibit prominent $C_2$-symmetric features, providing crucial evidence for the rotational symmetry breaking of the quasiparticle mass enhancement.

In the context of Kondo hybridization, heavy electrons carrying a magnetic moment can interact with and spin-polarize the conduction electrons. To gain magnetic information of the nematic order, we have conducted spin polarized STM. In the experiment, we used a Nickel tip to probe the sample[54-56]. For tunneling junctions between magnetic materials, due to matrix element effects, the tunneling current and conductance is enhanced when the spin polarizations of the tip and sample are aligned and suppressed when they are anti-aligned. Nickel is a soft ferromagnet whose spin polarization can be tuned by applying a small magnetic field[54-56], whereas the bulk USbTe crystal has a coercive field B ~ 1.3T at 5K (see Extended Fig. 8). Thus, by applying fields along the c-axis of +0.5T and -0.5T, we can change the tip polarization without affecting the sample magnetization (shown schematically in Fig. 4**c**), allowing us to probe the sample's spin-polarized states. Figure 4**a** shows the area of our focus taken on the U termination, where individual U atoms are marked. We first applied a magnetic field of +4T (up) to polarize the sample and tip and ramped it down to 0T. Then, we measured the tunneling conductance under +0.5T (up) and -0.5T (down) on the same linecut as marked, i.e. along the diagonal direction of the U lattice. To make sure data are taken at the same location, we find a reference impurity across the linecut (black marker). While the magnetic field is reversed from +0.5T to -0.5T, the topographic height, representing the atomic corrugation of the U atoms, stays the same (Fig. 4**b**). However, the tunneling conductance dI/dV at the Fermi level exhibits a contrast reversal (Fig. 4**b**): while at +0.5T the tunneling conductance peaks at the U site (red curve), at -0.5T the signal turns to a local minimum at the same atomic location (blue curve). The contrast reversal is quite robust near the Fermi level and can be well reproduced with different Ni tips and sample [Extended Fig.5]. To rule out accidental tip change, we have applied a sequence of magnetic fields +0.5T → -0.5T → +0.5T → -0.5T. The contrast reversal appears as long as the field is flipped, as shown in Fig. 4**d**. To confirm that the signal reversal comes from spin channels, we repeat the measurement with non-polarized Pt-Ir tip and show the results in Extended figure 3. With the Pt-Ir tip, we do not observe any contrast reversal under ±0.5T. The result shows that during the field flipping, the sample does not experience noticeable changes of electronic states, further supporting the conclusion that the observed contrast reversal comes from the spin sensitivity of the tip. The results indicate a spin polarization of the heavy fermion lattice: Local moments from f-electrons sit at the U sites and have the same polarization as the sample (up). At +0.5T, the states at the U sites (up) have the same spin polarization as the tip (up), which enhances the tunneling process. At -0.5T, the states at the U sites (up) have opposite spin polarization as the tip (down), suppressing the tunneling process. Importantly, between the U sites, the tunneling signal at -0.5T surpasses that at 0.5T, providing signatures of an antiferromagnetic type of spin density wave.

We move to investigate the nematic state using the spin-polarized tip by performing zoomed-in dI/dV maps under ±0.5T. Figure 4**e** shows the topographic images and the corresponding dI/dV maps. The marked impurity is referenced to ensure that the two maps are taken in the same area. Under +0.5T, the tip with up polarization (aligned with the sample) visualizes nematic stripes along the vertical direction (45 degrees to *a* and *b* axes). However, under -0.5T, the tip with down polarization (anti-aligned with the sample) does not yield a nematic pattern. To quantitatively validate our results, we have conducted a set of FT-dI/dV maps (see Extended Fig.6) and calculated the energy-dependent α values. Figure 4**f** plots the α value for the FT-dI/dV maps under ±0.5T. At +0.5T, α peaks near 0 meV, similarly to the non-spin-polarized result, indicative of the dominant symmetry-broken states at the Fermi level. In contrast, at -0.5T (tip polarization is flipped), the peak is substantially weakened. To rule out that the applied magnetic field changes the nematic state, we have conducted field-dependent maps with non-polarized Pt-Ir tips. As shown in the Extended Fig.5, focusing on the same area, the stripe pattern does not exhibit field dependence. Therefore, the sensitivity of α to the tip's polarization reveals that the symmetry-broken states tend to tunnel through a particular spin channel. The qualitative spectroscopic patterns and quantitative α values provide direct evidence that the nematic state has spin polarization aligned with the local moment. Whether the nematic state would break time-reversal symmetry on its own or whether the spin-polarization is a by-product of the ferromagnetic transition at much higher temperatures cannot be distinguished from our data.

We now discuss the interpretation of the experimental data. USbTe has a tetragonal structure with *P4/nmm* space group and $D_{4h}$ point group. In the nematic phase, the tetragonal symmetry is broken and the diagonals on the (*a,b*) plane become inequivalent. This is indicative of a nematic order parameter that transforms as the $B_{2g}$ irreducible representation (irrep) of the tetragonal group (i.e., $d_{xy}$ form-factor symmetry); incidentally, this is analogous to the nematic order observed in iron pnictides, which also have the same space group *P4/nmm*[8]. In our work, we observe that the nematic order is substantially suppressed above 15K. Importantly, at a similar temperature range, we observe a peak-dip feature in the dI/dV spectra, as well as the formation of a heavy band in QPI below 15K. Similar STM features, but without tetragonal symmetry breaking, were also observed in the heavy-fermion $URu_2Si_2$ around 17K, where they were attributed to band hybridization and the onset of a hidden order state[51,57]. Interestingly, in both systems, electrical transport measurements suggest a higher Kondo temperature[30,37], at which Kondo hybridization is generally believed to take effect. The reason why STM observes the formation of heavy quasi-particles at a lower temperature deserves further studies. Notwithstanding this caveat, our data clearly reveals tetragonal symmetry breaking, and thus nematic order, accompanied by the formation of heavy quasiparticles. The electronic nature of nematicity is supported by several observations. First, the topographic images reveal the lack of significant structural disorder in the temperature range where nematicity is seen (Extended Fig.1 d), and no detectable lattice distortion. By examining the topography across a nematic domain (Extended Fig. 1c), we do not observe structural changes. Second, we observe nematic domains, which are characteristic of long-range order rather than local order induced by extrinsic global strain, similarly to what was observed previously in nematic iron pnictides[44]. Third, the spin polarized STM data shows that the nematic order occurs predominantly on a particular spin channel, which supports an inherently electronic origin for this phenomenon. Another important question about the observed nematicity is whether it is a bulk feature or comes from surface reconstruction. We note that the nematic order is observed on both Uranium and Antimony terminations [Extended Fig.9] and does not adhere to a specific kind of surface. While simultaneously mapping the topography and conductance maps, no structural changes are detected across the nematic domain [Extended Fig.1]. Therefore, we conclude that nematicity does not come from structural surface effects. Moreover, below 15K, bulk electrical transport measurement on USbTe reports a resistivity upturn[37]. This is consistent with our STM measurements: (a) Below 15K, we observe a dip in dI/dV spectra near the Fermi level, signaling a suppression of the density of states. (b) The concurrent emergence of a nematic stripe order that can cause additional scattering.

Since nematic order concurs with the formation of heavy quasi-particles, it is natural to consider a scenario in which the Kondo hybridization spontaneously breaks the tetragonal symmetry of the system. To further explore this scenario, we consider the hybridization order parameter $\Delta = \langle \sum f_{k\sigma}^{\dagger} c_{k\sigma} + \text{h.c.} \rangle$ where $c_{k\sigma}$ is the fermionic operator for the itinerant electrons (which can be the *d*-orbitals of U or the *p*-orbitals of Sb) and $f_{k\sigma}$ refers to the local *f*-orbitals of the U atom. As discussed above, the symmetry of the nematic state extracted from QPI is that of a nematic order parameter transforming as the $B_{2g}$ irrep of the point group $D_{4h}$. A straightforward group-theory classification of the possible hybridization order parameters involving the *p,d,f*-orbitals [Extended Table 1,2] reveals that the *p-f* hybridization provides two routes to realize $B_{2g}$ nematic order: either as a hybridization between $f_{xyz}$ and $p_z$ or as a hybridization between $(f_{xz^2}, f_{yz^2})$ or $(f_{x(x^2-3y^2)}, f_{y(3x^2-y^2)})$ and $(p_x, p_y)$. In contrast, since the *d* and *f* orbitals have opposite parities, their combination is always odd under inversion and thus cannot give a pure nematic order. We note, however, that *d-f* hybridization can result in an order parameter that transforms as the $E_u$ irrep, which, besides breaking inversion symmetry, can also break the tetragonal symmetry via a secondary composite order. Since the QPI map is always centrosymmetric, it has limitations to fully capture the symmetry properties of the Fermi surface under inversion. Additional work is thus needed to completely rule out such a scenario. Interestingly, the mechanism in which the *p-f* hybridization is responsible for the nematic order is consistent with the fact that the conduction band seen in the ARPES measurement disperses steeply at a large energy range (Fig.3). Such a steep dispersion is indicative of a weak mass renormalization, which in turn is consistent with a *p*-orbital origin. It is important to emphasize that a non-zero symmetry-breaking hybridization order parameter will only result in an anisotropic electronic spectrum in the presence also of a symmetry-preserving hybridization, as discussed in methods. In this regard, it is interesting to note the higher Kondo temperature estimated from transport, which may be related to the onset of such a symmetry-preserving hybridization.

Our work therefore reveals a direct relationship between Kondo hybridization and nematic order in USbTe. While the phenomenological scenario proposed here is consistent with several experimental observations, further work is needed to develop a microscopic model for the interplay between different types of symmetry-breaking and symmetry-preserving hybridization order parameters. Our findings also raise the question of whether a similar effect could address the in-plane anisotropic responses previously seen in other heavy fermions, such as CeRhIn$_5$[31] and URu$_2$Si$_2$[32]. More broadly, the mechanism of hybridization-driven nematicity might also be applicable to other quantum materials such as nematic order in twisted bilayer graphene[58,59].

**Acknowledgement:** M.Z.H. acknowledges support from the US Department of Energy, Office of Science, National Quantum Information Science Research Centers, Quantum Science Center and Princeton University. M.Z.H. acknowledges visiting scientist support at Berkeley Lab (Lawrence Berkeley National Laboratory) during the early phases of this work. Theoretical and STM works at Princeton University were supported by the Gordon and Betty Moore Foundation (GBMF9461; GBMF4547; M.Z.H.). R.M.F. was supported by the U.S. Department of Energy, Office of Science, Basic Energy Sciences, Materials Science and Engineering Division, under Award No. DE-SC0020045. E.M. acknowledges the support of Fapesp (Grant No. 22/15453-0) and CNPq (Grant No. 309584/2021-3).

**Competing interest:** The authors declare no competing interests.

**Data availability:** The datasets supporting the results of this study are available from the corresponding authors upon reasonable request.

## Methods

**Crystal growth:**

Single crystals of USbTe were synthesized by the chemical vapor transport method using iodine as the transport agent. Elements of U, Sb, and Te with atomic ratio 1:0.8:0.8 were sealed in an evacuated quartz tube, together with 1 mg/cm3 iodine. The ampoule was gradually heated up and held in the temperature gradient of 1030/970 ∘C for 7 days, after which it was furnace cooled to room temperature.

**STM experiments:**

USbTe single crystals were cleaved at 78K under ultrahigh vacuum (UHV) conditions ($< 5 \times 10^{-10}$mbar), and then immediately inserted into the microscope head, which is at the $^4$He base temperature (4.2 K). The cleaved sample shows a [001] surface with atomic flatness. Commercial STM Pt-Ir tip (nonmagnetic) and STM Nickel tip (soft magnet) were used in this study. Tunneling conductance maps were obtained using standard lock-in amplifier techniques with a lock-in frequency of 974 Hz and tunneling junction set-ups as indicated in the corresponding figure captions. In temperature-dependent measurement, STM tip was first retracted by several steps from the sample before the heat source was applied. Data was taken after the temperature was stabilized and the STM system was relaxed, which usually takes hours. QPI maps in Fig.3 are two-fold symmetrized and smoothed using Gaussian window functions and raw data is provided in Extended Fig. 2. In this study, we have successfully 5 samples and observed nematicity on all the cleaved samples.

For magnetic field dependent measurements, data was taken while the superconducting magnet is in persistent mode. While ramping the field, we carefully retracted the STM tip from the sample to avoid tip crashing. We have used the reference defect to make sure dI/dV maps at different fields are taken at the same area. Compared to dI/dV maps, it is more difficult to make sure the dI/dV linecuts (without crossing any defect, as shown in Fig. 4d and Extended Fig.4) are taken at the same area. To make sure the linecuts are taken at the same position, we have taken the following procedure: After changing the magnetic field, we first find the same region using the reference defect. Then the tip is relaxed to avoid drifting. The relaxation can take hours. During the relaxation, we repeatedly scan zoom-in topography image to correct the drift. After the procedure, we take the linecut at the desired position. After the linecut, we scan the topography again to make sure the drift is negligible.

**Calculating symmetry factor α:**

In the work, we have introduced symmetry factor α to quantitatively compare the rotational symmetry breaking of FFT images. The procedure for calculating α is as follows: We first apply discrete Fourier transform to real space dI/dV maps and get a set of FFT images. We then apply a constant window to screen out Bragg peaks, area outside the Bragg peaks, and the noise at the FFT image center. The FFT image is then coded by a 2D matrix. We then renormalized the matrix to get M and $M^R$. Finally, we calculate $\alpha = Tr[(M - M^R)(M - M^R)^T]$.

To accurately reflect the symmetry of the raw data, we did not apply any symmetrization to the FFT images for calculating α. For a perfect four-fold rotational symmetric matrix, α = 0. In the experiment, due to unavoidable noise, α always has a finite value. Therefore, we focus on the trend of α as a function of energy, temperature, and spin-sensitivity, rather than its absolute value.

**ARPES measurement:**

Angle-resolved photoemission spectroscopy experiments were performed at the beamline 5-2, Stanford Synchrotron Radiation Lightsource in Menlo Park, USA. Samples were cleaved in the ultra-high vaccum (<10^11mbar) at 130K and gradually cooled down to the base temperature. Unless otherise stated, the impinging light is p-polarized with an energy of 98 eV. The Fermi levels of the energy-momentum cuts were determined by measuring polycrystalline gold under the same condition. The energy resolution of cuts is better than 10 meV.

**Theoretical model:**

Here we propose a simple phenomenological model to describe nematic hybridization in USbTe. This material has space group *P4/nmm* and thus $D_{4h}$ tetragonal point group. The STM data reveals tetragonal symmetry breaking along the diagonals on the *(a,b)* plane. This is consistent with a nematic order with a form factor $k_x k_y$ in momentum space, which transforms as the $B_{2g}$ irreducible representation (irrep) of the point group $D_{4h}$ and lowers the point group to the orthorhombic $D_{2h}$.

We first use group theory to determine which type of hybridization order parameter $\Delta = \langle \sum f^\dagger_{k\sigma} c_{k\sigma} +$ h.c.$\rangle$ transforms as the $B_{2g}$ irrep. Here, $c_{k\sigma}$ is the fermionic operator for the itinerant electrons and $f_{k\sigma}$ refers to the local $f$-orbitals of the U atom. There are seven different $f$ orbitals, which we label in terms of the usual cubic harmonics $f_{z^3}, f_{xz^2}, f_{yz^2}, f_{xyz}, f_{x(x^2-3y^2)}, f_{y(3x^2-y^2)}$, and $f_{z(x^2-y^2)}$. In the presence of the crystal fields of USbTe, they transform as 4 different irreps: $A_{2u}, B_{1u}, B_{2u}$ and $E_u$, as shown in the Extended Tables 1 and 2. As for the itinerant electrons $c_{k\sigma}$, we can consider either the U d-orbitals or the Sb p-orbitals. In the first case, there are five different orbitals, $d_{z^2}, d_{x^2-y^2}, d_{xy}, d_{xz}, d_{yz}$, which transform as 4 different irreps: $A_{1g}, B_{1g}, B_{2g}$ and $E_g$ (see Extended Table 1). In the latter case, there are three different orbitals $p_x, p_y, p_z$ which transform as 2 irreps: $A_{2u}$ and $E_u$ (see Extended Table 2).

In Extended Tables 1 and 2, we show the symmetry of the hybridization order parameter for all possible combinations between $f$ and $d$ or $p$ orbitals in terms of the irreps of $D_{4h}$. If we enforce the condition that the hybridization order parameter must be purely nematic and transform as $B_{2g}$, there are only three options:

$$\Delta = \langle \sum_{k\sigma} f^\dagger_{xyz, k\sigma} p_{z, k\sigma} + h.c. \rangle \tag{1}$$

$$\Delta = \langle \sum_{k\sigma} f^\dagger_{xz^2, k\sigma} p_{y, k\sigma} + f^\dagger_{yz^2, k\sigma} p_{x, k\sigma} + h.c. \rangle \tag{2}$$

$$\Delta = \langle \sum_{k\sigma} f^\dagger_{x(x^2-3y^2), k\sigma} p_{y, k\sigma} - f^\dagger_{y(3x^2-y^2), k\sigma} p_{x, k\sigma} + h.c. \rangle \tag{3}$$

The fact that none of the combinations involving $f$-$d$ hybridization gives a pure nematic order parameter is a simple consequence of the fact that the two orbitals have opposite parities. However, if we relax the condition that the order parameter must be a pure $B_{2g}$ nematic, but allow it to break other point-group symmetries besides fourfold rotations, the number of options increases substantially. In this case, $\Delta$ transforming as $E_u$ or $E_g$ can also support a secondary composite $B_{2g}$ nematic order parameter. To understand this, we note that because $E_u$ and $E_g$ are two-dimensional irreps, the hybridization gap is a two-component order parameter $\Delta = (\Delta_1, \Delta_2)$. It turns out that we can combine these two components to define a composite order parameter $\varphi = \Delta_1 \Delta_2$ that transforms precisely as the $B_{2g}$ nematic order parameter. It is important to emphasize that if either an $E_u$ or an $E_g$ type of hybridization order parameter $\Delta$ condenses, other symmetries of the tetragonal lattice will be broken besides the fourfold rotational symmetry. In particular, if only fourfold rotations are broken by the hybridization (i.e. $\Delta$ transforms as $B_{2g}$), the resulting point group is the orthorhombic $D_{2h}$, which possesses inversion and two vertical mirrors. In contrast, if the condensed $\Delta$ transforms as $E_g$, the resulting point group is the monoclinic $C_{2h}$, which possesses inversion but no vertical mirrors. Finally, if $\Delta$ transforms as $E_u$, its condensation lowers the point group to the orthorhombic $C_{2v}$, which does not possess inversion although it does have vertical mirrors. Further experiments are needed to unambiguously determine whether the system has vertical mirrors and inversion in the nematic phase.

Having established that only hybridization order parameters of the form Eq (1)-(3) transform as a pure nematic $B_{2g}$ order parameter, we now discuss its fingerprints on the electronic dispersion of the heavy Fermi liquid state by using a standard mean-field description[60-64]. We start by considering the case of $p_z$ itinerant electrons and $f_{xyz}$ local orbitals, with hybridization given by Eq (1). The mean-field Hamiltonian is given by $H = H_0 + H_{\text{hyb}}$ with:

$$H_0 = \sum_{k\sigma} \varepsilon_k p_{k\sigma}^\dagger p_{k\sigma} + \tilde{E}_f \sum_k f_{k\sigma}^\dagger f_{k\sigma}$$

$$H_{\text{hyb}} = \sum_{k\sigma} \tilde{\Delta}_k (f_{k\sigma}^\dagger p_{k\sigma} + p_{k\sigma}^\dagger f_{k\sigma}) + \sum_{k\sigma} \Delta (f_{k\sigma}^\dagger p_{k\sigma} + p_{k\sigma}^\dagger f_{k\sigma}) \quad (4)$$

where $\Delta$ is the hybridization order parameter given by Eq (1) and $\tilde{\Delta}_k = \tilde{\Delta} \sin k_x \sin k_y$ is the symmetry-preserving hybridization between $p$ and $f$ orbitals. Here, $\varepsilon_k$ is the dispersion of the itinerant electrons and $\tilde{E}_f$ is the renormalized f-electron energy. We would like to emphasize the difference between $\Delta$ and $\tilde{\Delta}_k$. The first one is the direct hybridization between $p$ and $f$ orbitals and as such it breaks the tetragonal symmetry of the lattice. Consequently, its onset must be accompanied by a phase transition. As for the second term, $\tilde{\Delta}_k$, it does not break any symmetry of the lattice, and as such it is non-zero at any temperature. In a heavy Fermi liquid description, $\tilde{\Delta}_k$ is expected to show a pronounced enhancement at the Kondo coherence temperature, displaying a crossover behavior rather than a phase transition.

To break the fourfold rotational symmetry, both $\Delta$ and $\tilde{\Delta}_k$ must be non-zero. To see this, we can integrate out the $f$ electrons and treat $H_{hyb}$ perturbatively in powers of $\Delta^2/\tilde{E}_f$ we find the following correction to the itinerant electron dispersion:

$$\delta H_0 = \frac{1}{\tilde{E}_f} \sum_{k\sigma} (\Delta^2 + \tilde{\Delta}^2 \sin^2 k_x \sin^2 k_y + 2\Delta\tilde{\Delta} \sin k_x \sin k_y) p_{k\sigma}^\dagger p_{k\sigma} \quad (5)$$

The first two terms preserve tetragonal symmetry, whereas the second one breaks it, as it clearly transforms as the $B_{2g}$ irrep. Therefore, the band dispersion in the heavy Fermi liquid phase, $E_\pm(k)$, exhibits nematic features only when $\Delta, \tilde{\Delta} \neq 0$, with:

$$E_\pm(k) = \frac{\varepsilon_k + \tilde{E}_f}{2} \pm \sqrt{\left(\frac{\varepsilon_k - \tilde{E}_f}{2}\right)^2 + (\Delta + \tilde{\Delta}_k)^2} \quad (7)$$

To illustrate this behavior, we show in Extended Fig.7 the dispersion (7) of the heavy Fermi liquid phase along the two diagonals, which are related by a 90° rotation. The red curve corresponds to $\Delta \neq 0$ and $\tilde{\Delta} = 0$, whereas the blue curve corresponds to $\Delta \neq 0$ and $\tilde{\Delta} \neq 0$. Clearly, the fourfold rotational symmetry of the band dispersion is broken in the latter case only.

The same situation happens in the case where the $(p_x, p_y)$ itinerant orbitals hybridize with either $(f_{x(x^2-3y^2)}, f_{y(3x^2-y^2)})$ or $(f_{xz^2}, f_{yz^2})$ orbitals, corresponding to the hybridization order parameters given by Eqs. (2) and (3). For simplicity of notation, we denote the orbitals as $(p_1, p_2)$ and $(f_1, f_2)$. The Hamiltonian in this case is given by:

$$H_0 = \sum_{k\sigma} (A_k \tau_0^{ij} + B_k \tau_z^{ij} + C_k \tau_x^{ij}) p_{i,k\sigma}^\dagger p_{j,k\sigma} + E_f \sum_{k\sigma} (f_{1,k\sigma}^\dagger f_{1,k\sigma} + f_{2,k\sigma}^\dagger f_{2,k\sigma})$$

$$H_{\text{hyb}} = \sum_{k\sigma} (\Delta + \tilde{\Delta}_k)(f_{1,k\sigma}^\dagger p_{2,k\sigma} \pm f_{2,k\sigma}^\dagger p_{1,k\sigma} + \text{h.c.})$$

where the plus sign corresponds to the case of $(f_{xz^2}, f_{yz^2})$ orbitals whereas the minus sign corresponds to the case of $(f_{x(x^2-3y^2)}, f_{y(3x^2-y^2)})$ orbitals. Here, $\tau_\mu$ are Pauli matrices in $p$-orbital space. As in the previous case, $\Delta$ is the hybridization order parameter given by Eqs. (2) or (3), whereas $\widetilde{\Delta}_{\boldsymbol{k}} = \widetilde{\Delta}\sin k_x \sin k_y$ is the symmetry-preserving hybridization. Moreover, the terms

$$A_{\boldsymbol{k}} = -(t_\sigma + t_\pi)(\cos k_x + \cos k_y) - \mu$$

$$B_{\boldsymbol{k}} = -(t_\sigma - t_\pi)(\cos k_x - \cos k_y)$$

$$C_{\boldsymbol{k}} = -2t_d \sin k_x \sin k_y$$

correspond to $p_x, p_y$ orbitals with nearest- and next-nearest-neighbor hopping terms on the square lattice. In the limit $t_d = 0$ (nearest-neighbor hopping only), an analytical expression for the four dispersions in the heavy Fermi liquid phase is available:

$$E_\pm^\epsilon(\boldsymbol{k}) = \left(\frac{A_{\boldsymbol{k}} + \epsilon B_{\boldsymbol{k}} + \widetilde{E}_f}{2}\right) \pm \sqrt{\left(\frac{A_{\boldsymbol{k}} + \epsilon B_{\boldsymbol{k}} - \widetilde{E}_f}{2}\right)^2 + \left(\Delta + \widetilde{\Delta}_{\boldsymbol{k}}\right)^2}$$

with $\epsilon = \pm 1$. The case $t_d \neq 0$ can be solved numerically in a straightforward way. In either case, we find indeed that a nematic distortion of the band dispersion is only present when both $\Delta, \widetilde{\Delta} \neq 0$.

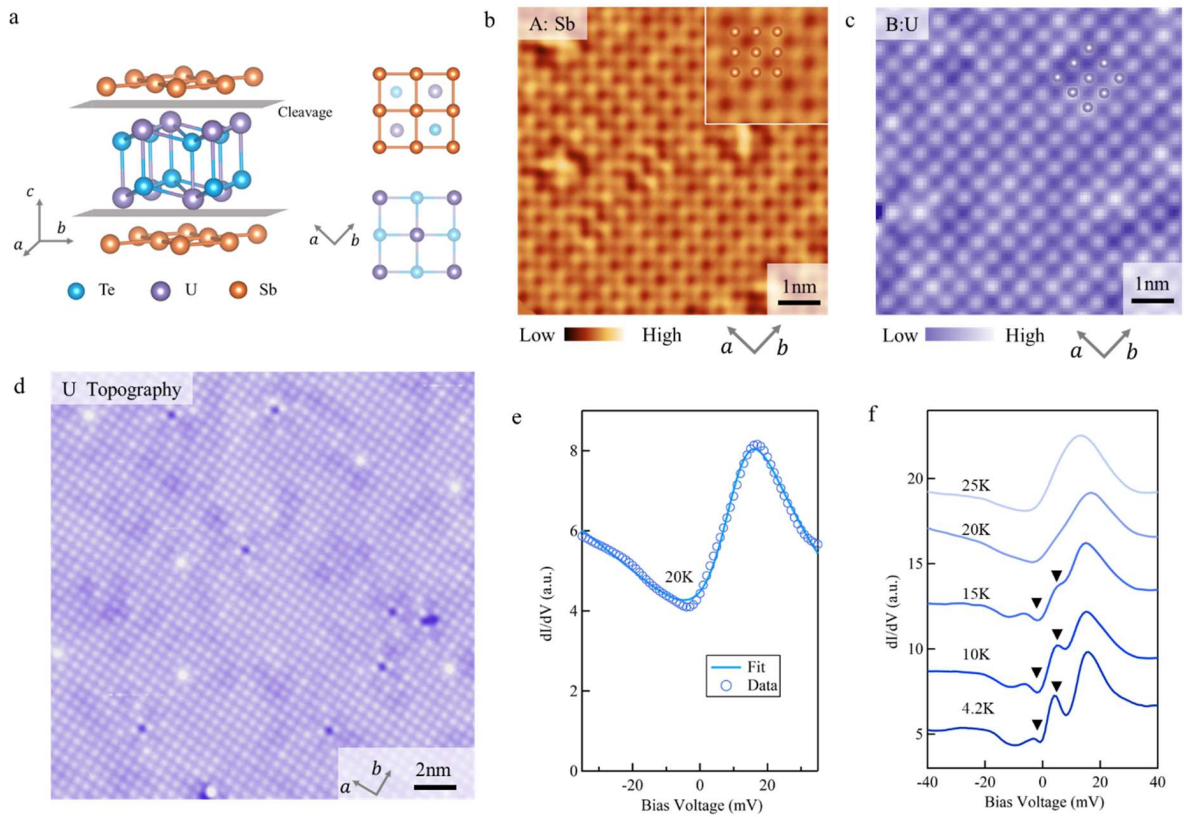

**Figure 1. Visualization of the Kondo lattice in USbTe. a,** Crystal structure of USbTe. The right panel shows the top view of the Sb and U planes with the underlying atoms. **b,c** Topography of Sb termination (left) and U termination (right). Atomic arrangements are labeled as a guide. Bias voltage: 100 mV. Set point current: 50 pA. **d,** Topography of the U termination showing atomic defects. **e,** dI/dV spectra on the U termination. Data is collected at 20K and fitted to a Fano spectrum with fitting parameters given in the main text. **f,** Temperature dependence of dI/dV spectra on the U termination. A constant offset is applied on each dI/dV spectra. At 15K, there is a peak-dip feature near the Fermi level, marked by the black arrow. Bias voltage: 50 mV. Set point current: 400 pA.

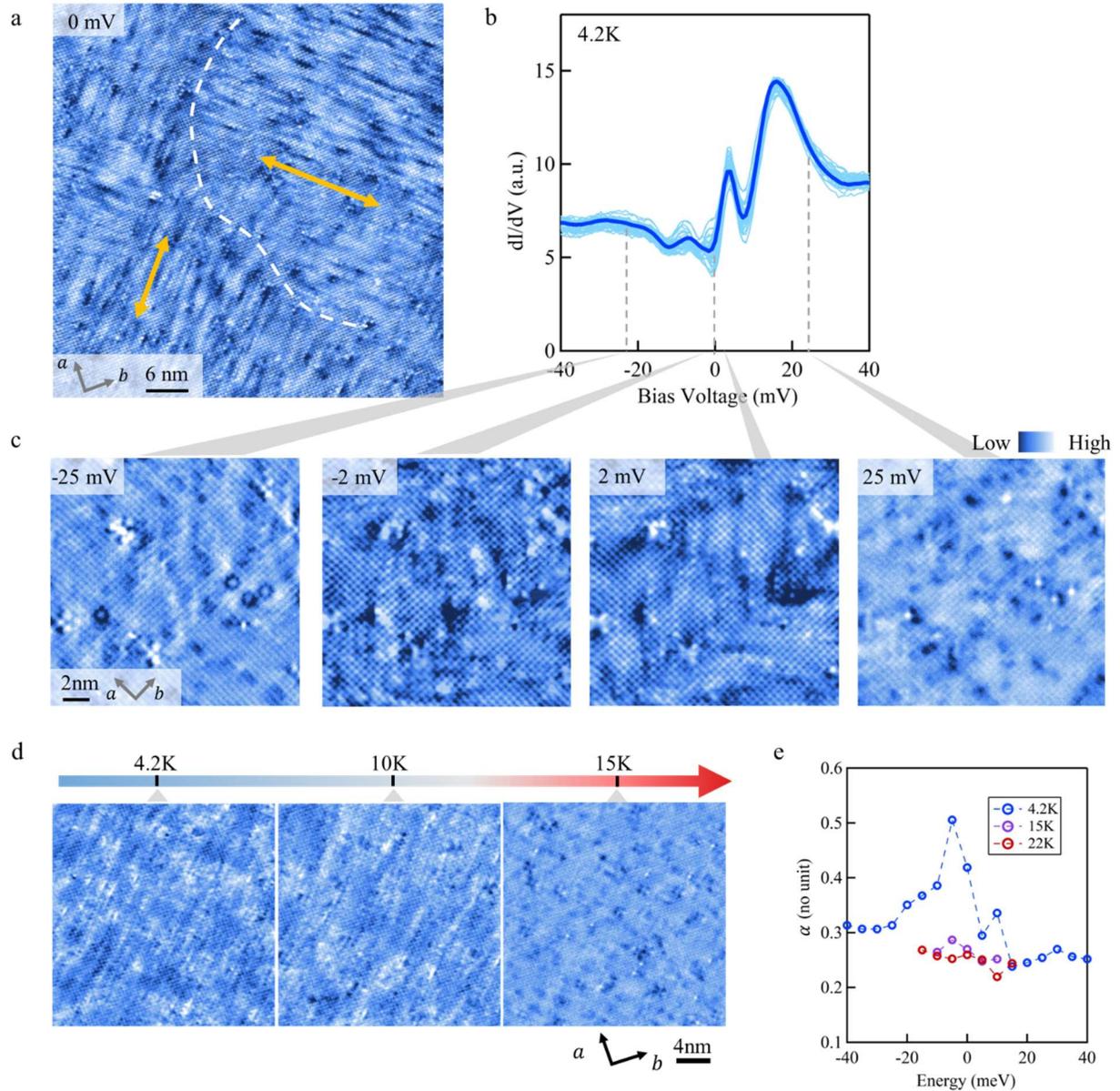

**Figure 2. Observation of a nematic Kondo lattice. a,** dI/dV map at 0 mV. Dashed line tracks a nematic domain wall. Yellow arrows mark the orientation of the anisotropic feature of the nematic state. **b,** dI/dV spectra on U termination. **c,** Energy-resolved dI/dV maps. Nematic order appears at ±2 mV. STM tunneling condition: V = 30 mV, I = 120 pA. The data in figure a-c are taken at 4.2K. **d,** dI/dV maps at 10 mV, taken at 4.2K, 10K, and 15K (left to right). **d,** Symmetry factor $\alpha$, which quantifies the four-fold rotational symmetry breaking. $\alpha$ peaks around the Fermi-level at 4.2K. Details on how the factor $\alpha$ is extracted are provided in the main text and Methods.

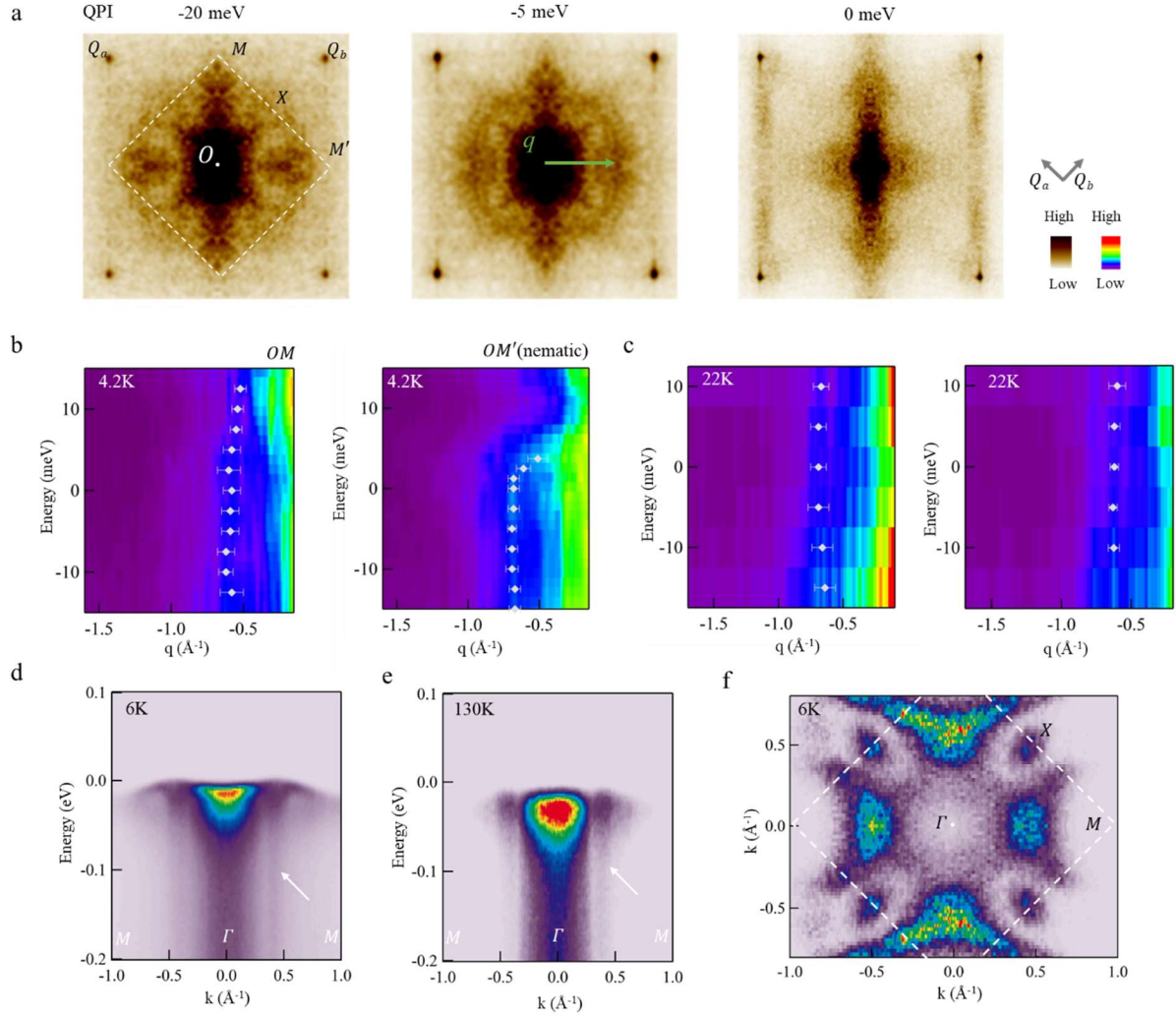

**Figure 3. Spectroscopic evidence of rotational symmetry breaking and heavy band formation. a,** Quasiparticle interference (QPI) maps taken at 4.2K. $Q_{a,b}$ labels the Bragg peaks. White dashed line corresponds to the first Brillouin zone, drawn for reference. The green arrow points to the scattering vector $q$. QPI patterns are two-fold symmetrized. **b,** Energy-momentum dispersion of QPI bands at 4.2K extracted along the $OM$ (left) and $OM'$ (right) directions. The white markers track the intensity bump at each energy cut. STM tunneling condition: $V_{bias}$ = 50 mV. I = 120 pA. $V_{mod}$ = 2.5 mV. **c,** Energy-momentum dispersion of QPI bands at 22K along $OM$ and $OM'$. **d,** ARPES-measured band dispersion along the $M - \Gamma - M$ direction at 6K. The white arrow points to a light band. **e,** Energy-momentum cut at 130K. **f,** ARPES-measured Fermi surface at $k_z$ = 0 at 6K. Data is two-fold symmetrized. ARPES measurement parameters are provided in methods.

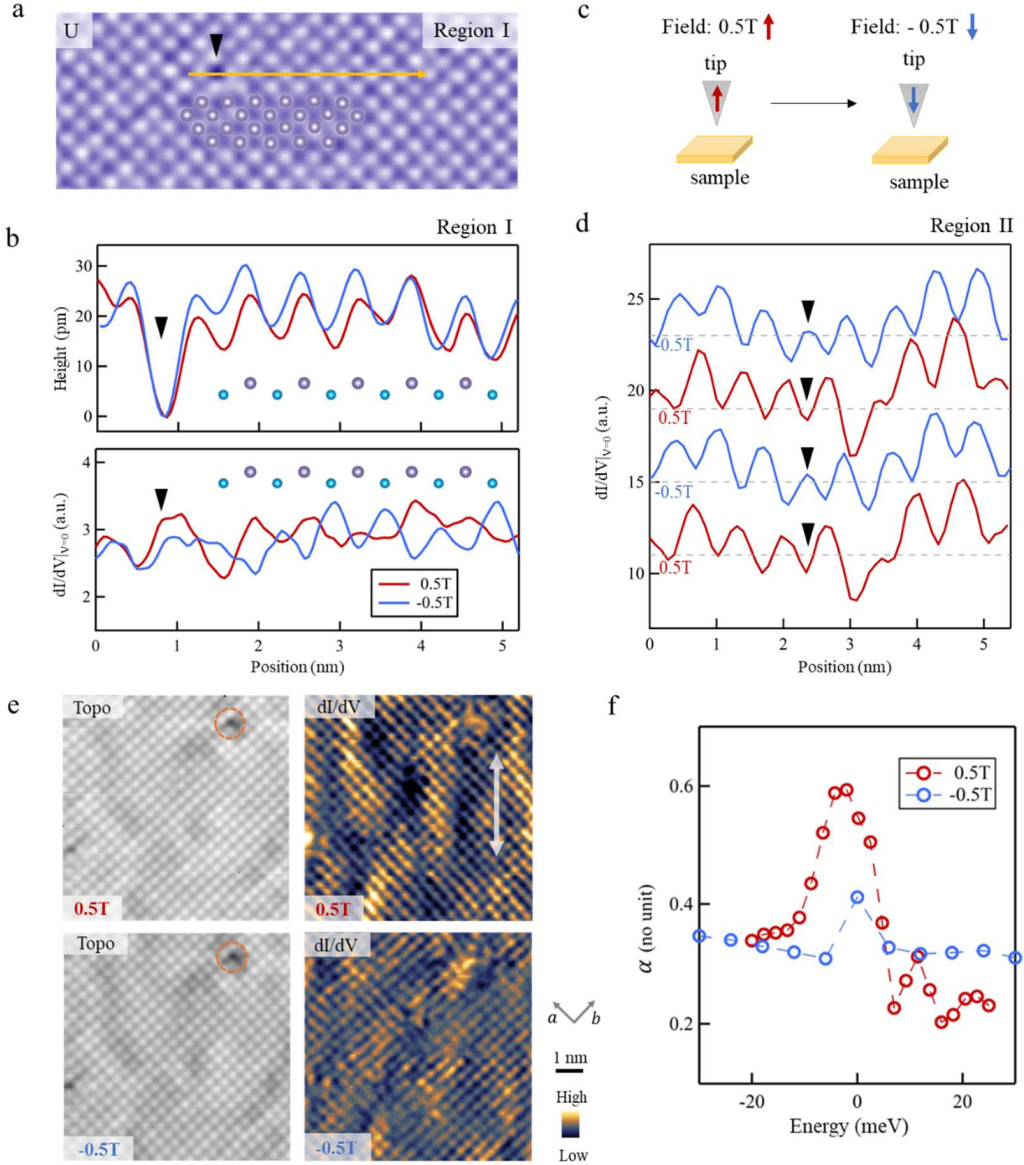

**Figure 4. Spin polarization of the electronic liquid crystal. a,** Topography of U termination acquired with Nickel tip in region I. The linecuts in figure b are extracted along the yellow line. The black marker corresponds to a reference impurity. **b,** Topographic corrugation (upper panel) and dI/dV(V=0) conductance (lower panel) along the linecut in region I at ±0.5T. The black marker points to the position of the reference impurity. STM tunneling conditions: V = 100 mV. I = 70 pA. **c,** Schematic of the spin polarized STM using Nickel tip. The blue and red arrows show the direction of the external field and the tip polarization, which is perpendicular to the sample's *ab* plane. **d,** dI/dV(V=0) conductance in region II under the sequence of fields +0.5T → -0.5T → +0.5T → -0.5T. An offset is applied, referenced by the dashed grey line. Topography images of region II are provided in Extended Fig. 4. **e,** Comparison of the topography and dI/dV (E = 3meV) maps in the same area. The reference impurity is marked. Upper and lower images are taken at +0.5T and -0.5T respectively. The white arrow guides the nematic stripe along 45 degrees to lattice vectors *a* and b, consistent with nematic modulations in Fig.2a. **f,** Symmetry factor $\alpha$ acquired with different tip polarizations.

18